\documentclass[11pt]{article}
\usepackage{moriond,epsfig}

  \def\Journal#1#2#3#4{{ #1} {\bf #2}, #3 (#4)}

  \def\NIM{\em Nucl. Instrum. Methods}

  \def\NP{\em Nucl. Phys.}
  
  \def\PL{\em Phys. Lett.}
  \def\PRL{\em Phys. Rev. Lett.}
  \def\PR{\em Phys. Rev.}

  \def\APP{\em Acta Phys. Polon.}

  \newcommand{\beq}[1]{\begin{equation}\label{#1}}
  \newcommand{\eeq}{\end{equation}}
  \newcommand{\bea}[1]{\begin{eqnarray}\label{#1}}
  \newcommand{\eea}{\end{eqnarray}}
  \newcommand{\Fi}[1]{Fig.~\ref{#1}}

  \newcommand{\De}{\Delta}
  \newcommand{\tm}{\ensuremath{\times}}
  \newcommand{\dedx}{\ensuremath{{\rm d}E/{\rm d}x}}
  \newcommand{\agev}{\ensuremath{A{\rm GeV}}}
  \newcommand{\mt}{\ensuremath{m_{\rm T}}}
  \newcommand{\bt}{\ensuremath{\beta_{\rm T}}}
  
  \newcommand{\kplus}{\ensuremath{{\rm K}^+}}
  \newcommand{\kmin}{\ensuremath{{\rm K}^-}}
  \newcommand{\pbar}{\ensuremath{\bar{\rm p}}}
  \newcommand{\nw}{\ensuremath{N_{\rm w}}}
  \newcommand{\kpiplus}{\ensuremath{{\rm K}^+ / \pi^+}}
  \newcommand{\kpimin}{\ensuremath{{\rm K}^- / \pi^-}}
  \newcommand{\lpiavg}{\ensuremath{\Lambda / \pi}}
  
  \newcommand{\tch}{\ensuremath{T_{\rm ch}}}
  \newcommand{\mub}{\ensuremath{\mu_{\rm B}}}

\begin{document}

%--------------------------------------------------------------------

% \vspace*{4cm} % for proceedings
\vspace*{2cm}   % for archive (to fit into 4 pages)

\title{NA49 ENERGY SCAN RESULTS FOR CENTRAL LEAD-LEAD COLLISIONS AT
  THE CERN SPS}

\author{ M. BOTJE for the NA49 Collaboration~\footnote{Presented at
    the XXXIX$^{\rm th}$ Rencontres de Moriond, La Thuile, March
    28--April 4, 2004.}
    }

\address{NIKHEF, PO box 41882,\\
1009DB Amsterdam, the Netherlands}

%--------------------------------------------------------------------

\maketitle\abstracts{
The energy dependence of hadron production in central Pb-Pb collisions
at SPS energies is presented and compared with data at lower and
higher energies and with results from p-p interactions. It is observed
that there is little change in transverse activity in the SPS energy
range, that there is a steepening rate of increase of pion production
and that the \kpiplus\ ratio exhibits a sharp peak located at about
30~\agev. The $\Lambda/\pi$ ratio also shows a pronounced maximum
which is weaker in $\Xi/\pi$ and absent in $\Omega/\pi$.}

%--------------------------------------------------------------------

\section{Introduction}

The NA49 collaboration has recently completed the energy scan program
at the SPS providing data on (central) Pb-Pb collisions at 20, 30, 40,
80 and 158 \agev\ beam energy. The aim of this program is to search
for the onset of a phase transition to the Quark Gluon Plasma.  That
such a transition may indeed occur at the SPS follows from the energy
density estimate~\cite{ref:edens} of about 3~GeV/fm$^3$ which is well
above the critical density of about 1~GeV/fm$^3$ calculated in Lattice
QCD.~\cite{ref:lattice} Assuming that a phase transition takes place
at about 30~\agev\ the Statistical Model of the Early stage
(SMES)~\cite{ref:smes} predicts anomalies in the energy dependence of
particle production at the SPS. In this paper we present a selection
of results from the energy scan; we refer to~\cite{ref:qm2004} and
references therein for a more complete overview of recent NA49
results.

%--------------------------------------------------------------------

\section{Experiment}

The NA49 detector~\cite{ref:na49nimpaper} is a large acceptance
fixed-target hadron spectrometer. Tracking is performed by four
large-volume TPC's. Two of these are placed one behind the other
inside two super-conducting dipole magnets (vertex TPC's). The two
other (main) TPC's are placed downstream of the magnets left and right
of the beam line. These main TPC's are optimized for particle
identification through a measurement of the specific energy loss
(\dedx) with a relative resolution of about 4\%. The combined TPC's
provide an accurate measurement of the particle momenta with a
resolution in the range of $\De p/p^2 \approx \mbox{0.3--7} \tm
10^{-4}$~(GeV/$c$)$^{-1}$. A measurement of the time-of-flight with a
resolution of about 60~ps provides particle identification at
mid-rapidity.  Centrality selection is based on a measurement of the
energy deposited by the projectile spectator nucleons in a forward
calorimeter.

In the years 1996 and 2000 two samples (0.8M and 3M events) of central
Pb-Pb collisions were collected at 158~\agev\ beam energy with a
centrality selection of, respectively, 10\% and 20\% of the inelastic
cross-section. In addition, Pb-Pb data (all with 7\% centrality
selection) were taken in the years 1999--2002 at beam energies of 80
(0.3M~events), 40 (0.7M), 30 (0.4M) and 20~\agev\ (0.3M). Results of
the 158, 80 and 40~\agev\ runs are published in~\cite{ref:epap} while
those of 30 and 20~\agev\ are still preliminary.

%--------------------------------------------------------------------

\section{Transverse mass spectra}   

In the left-hand plots of \Fi{fig:blastwave}
%
%------------------------
\begin{figure}[tbh]
\begin{center}
\mbox{\epsfig{figure=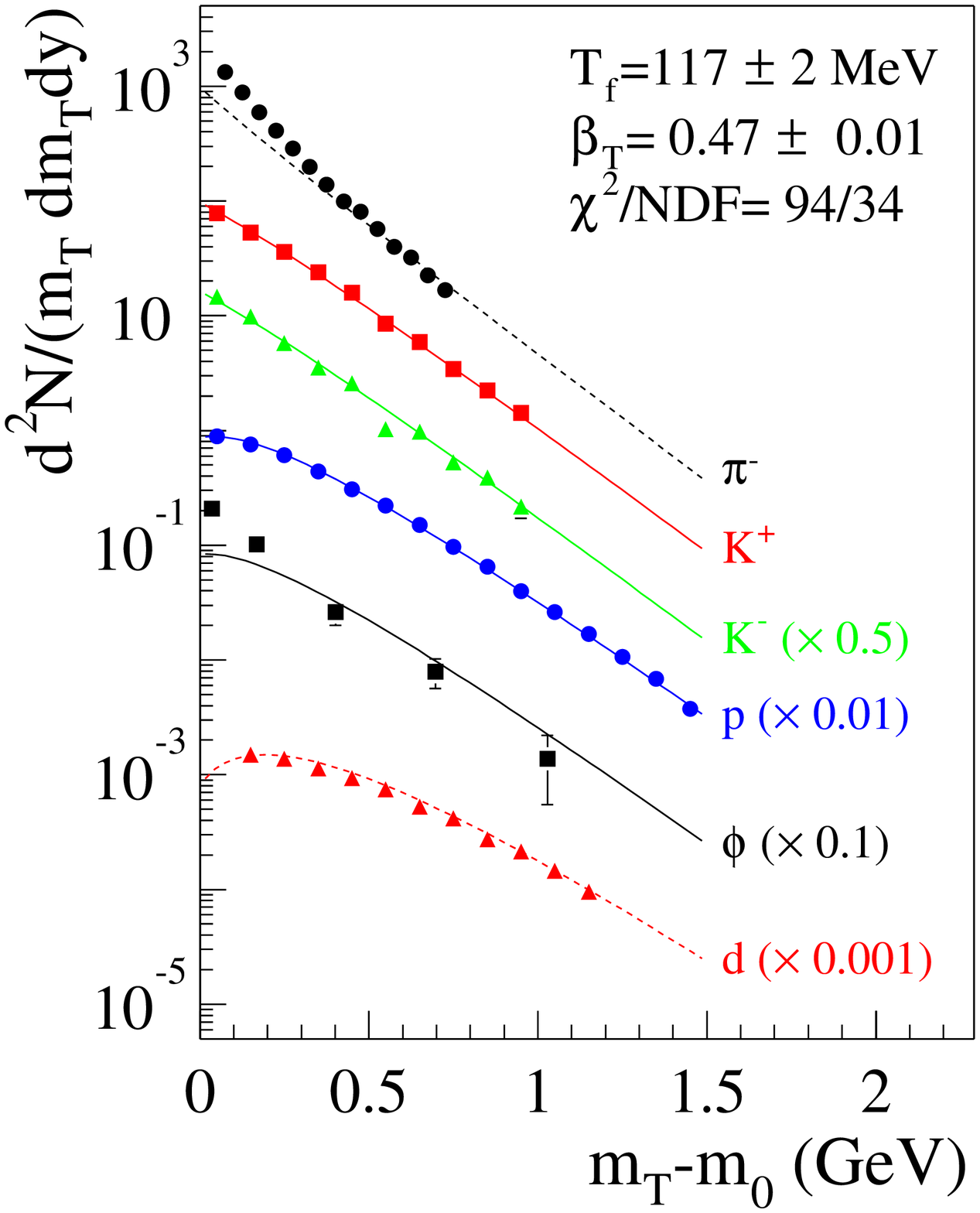,width=0.25\linewidth}
      \epsfig{figure=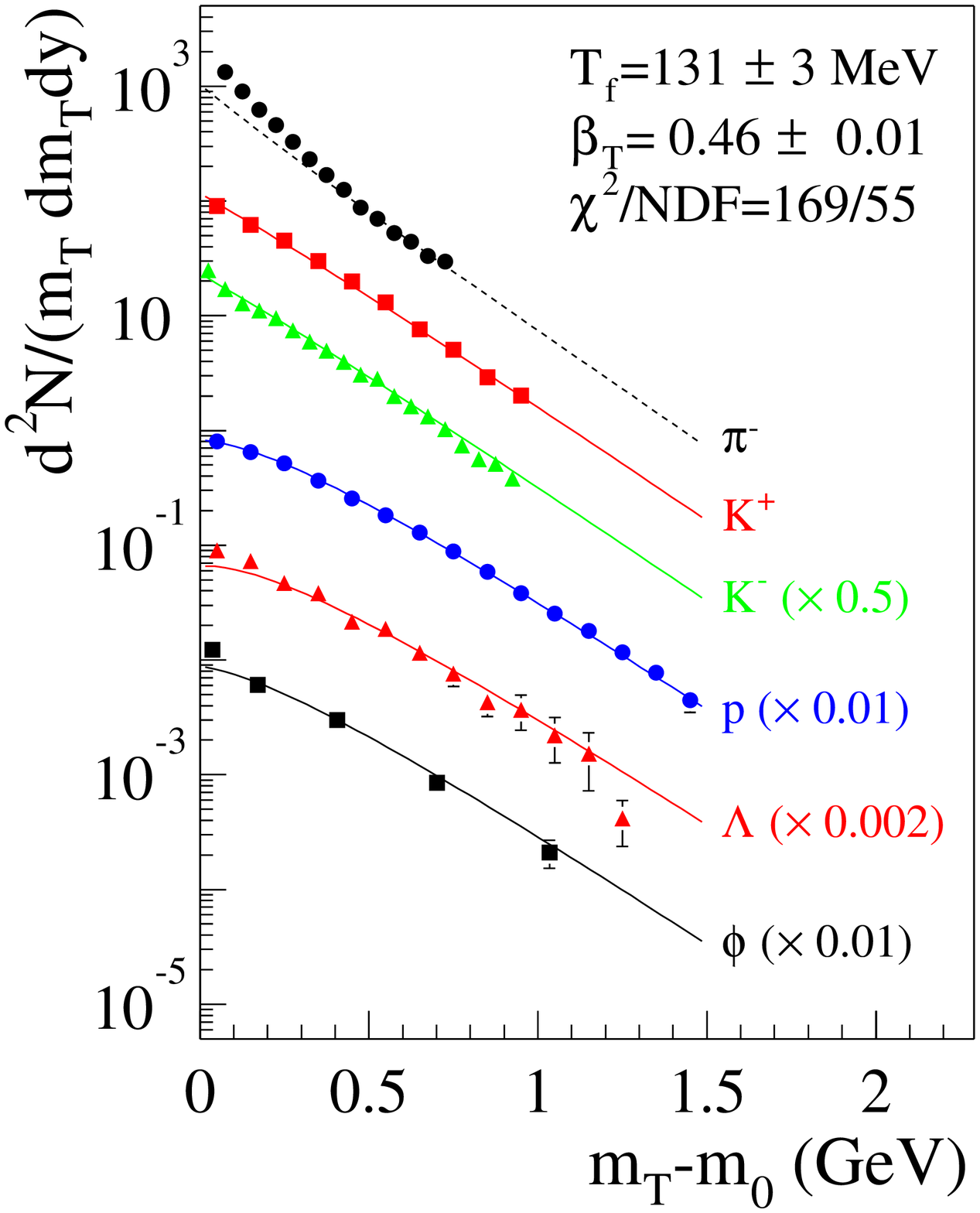,width=0.25\linewidth}
      \epsfig{figure=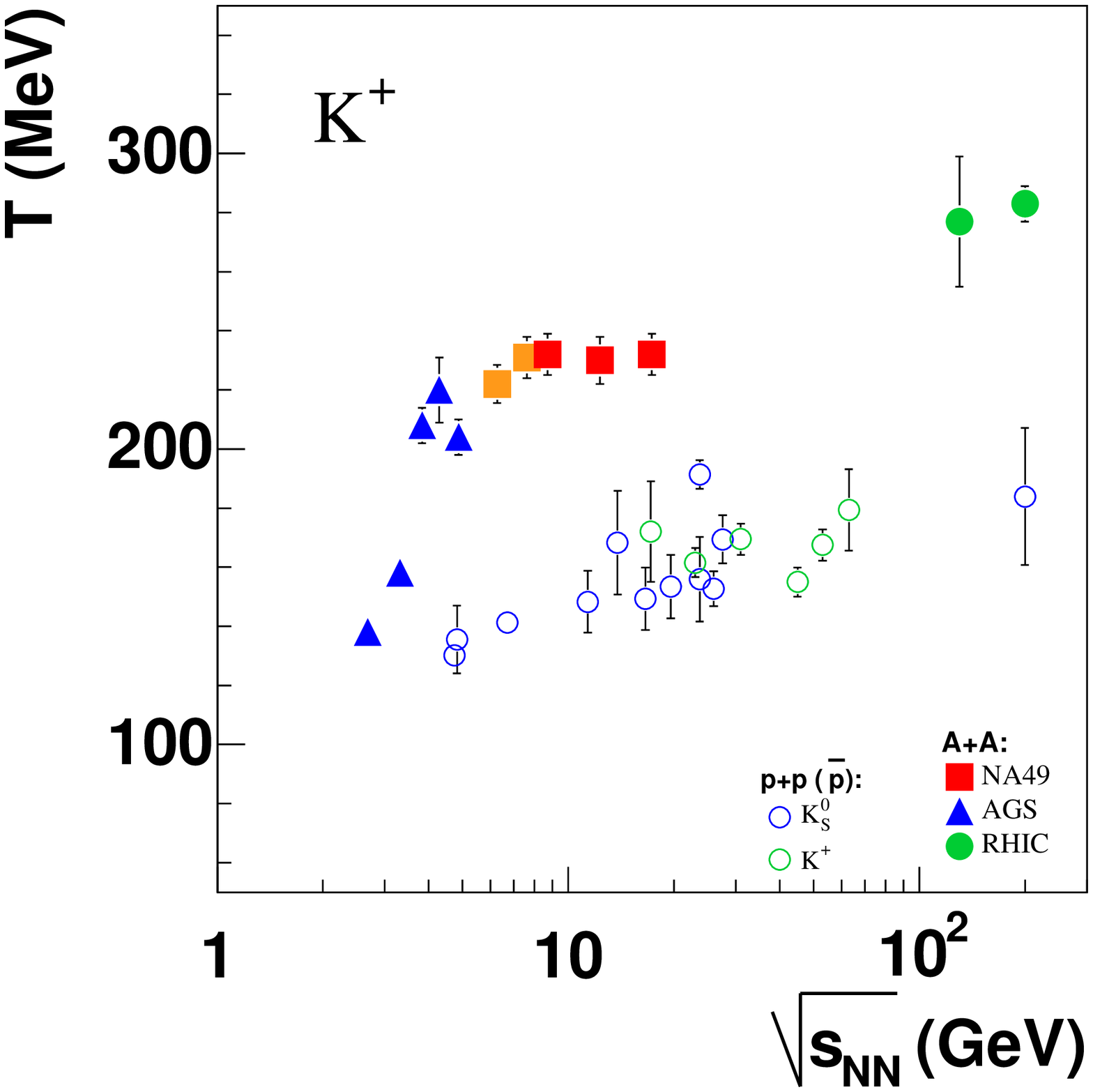,width=0.25\linewidth}
      \epsfig{figure=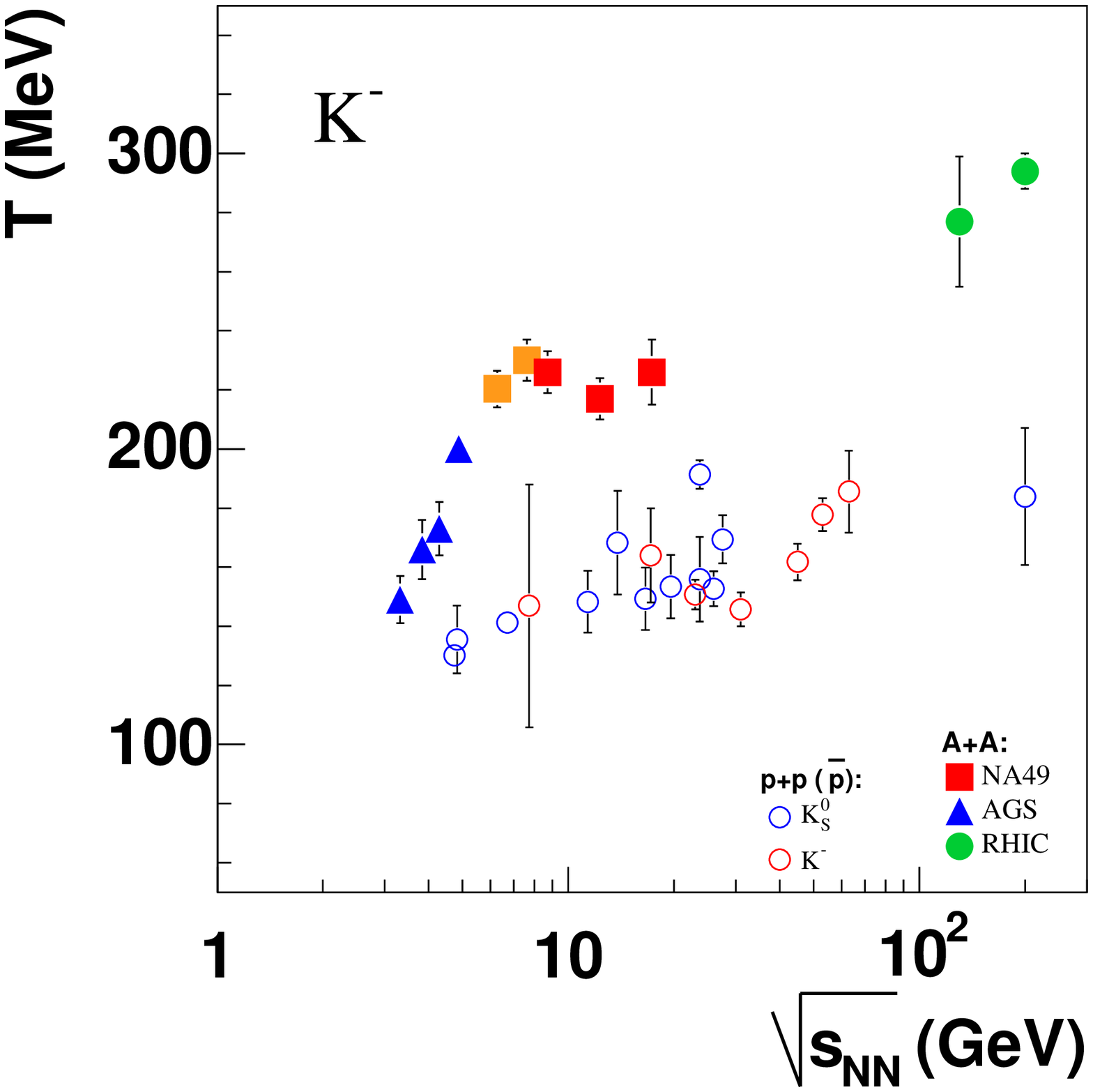,width=0.25\linewidth} }
\end{center}
\caption{Left: Transverse mass spectra of hadrons produced in central Pb-Pb
      collisions at 20 and 30~\agev. The lines
      correspond to a transverse flow fit (see text).
      Right: The inverse slope parameters ($T$) versus energy
      of \kplus\ and \kmin\ transverse mass
      spectra in central A-A collisions (full
      symbols) and in p-p interactions (open circles). 
\label{fig:blastwave}}
\end{figure}
%------------------------
%
are shown the transverse mass (\mt) spectra of hadrons measured at 20
and 30~\agev. The spectra are reasonably well described by a
blast-wave parameterization~\cite{ref:schneder} with a freeze-out
temperature $T \approx$ 120~MeV and a transverse flow velocity $\bt
\approx 0.5$. Similar values were found from the \mt\ spectra at
higher energies~\cite{ref:marco} indicating that there is little
change in transverse activity in the SPS energy range.  A similar
conclusion can be drawn from the right-hand plots of
\Fi{fig:blastwave} where we show the inverse slope parameters $T$
obtained from exponential fits to the \kplus\ and \kmin\ transverse
mass spectra measured at AGS, SPS and RHIC. It is seen from this
figure that for both \kplus\ and \kmin\ the inverse slope parameter
increases strongly with energy at the AGS, remains approximately
constant at the SPS and shows a tendency to increase further above SPS
energies. Such an energy dependence is in qualitative agreement with a
softening of the equation of state due to a phase transition at the
SPS.\cite{ref:hove,ref:step} The stationary value of the apparent
temperature $T$ would then indicate that the early stage
temperature and pressure remain constant due to the coexistence of
partonic and hadronic phases. The p-p(\pbar) data (open circles in
\Fi{fig:blastwave}) show an increasing trend but are not accurate
enough to reveal any structure.

%--------------------------------------------------------------------

\section{Energy dependence of particle yields}   

The rapidity distributions of $\pi^-$, \kplus\ and \kmin\ measured at
all energies are shown in the left-hand plot of \Fi{fig:rapidity}.
%
%------------------------
\begin{figure}[tbh]
\begin{center}
\mbox{\epsfig{figure=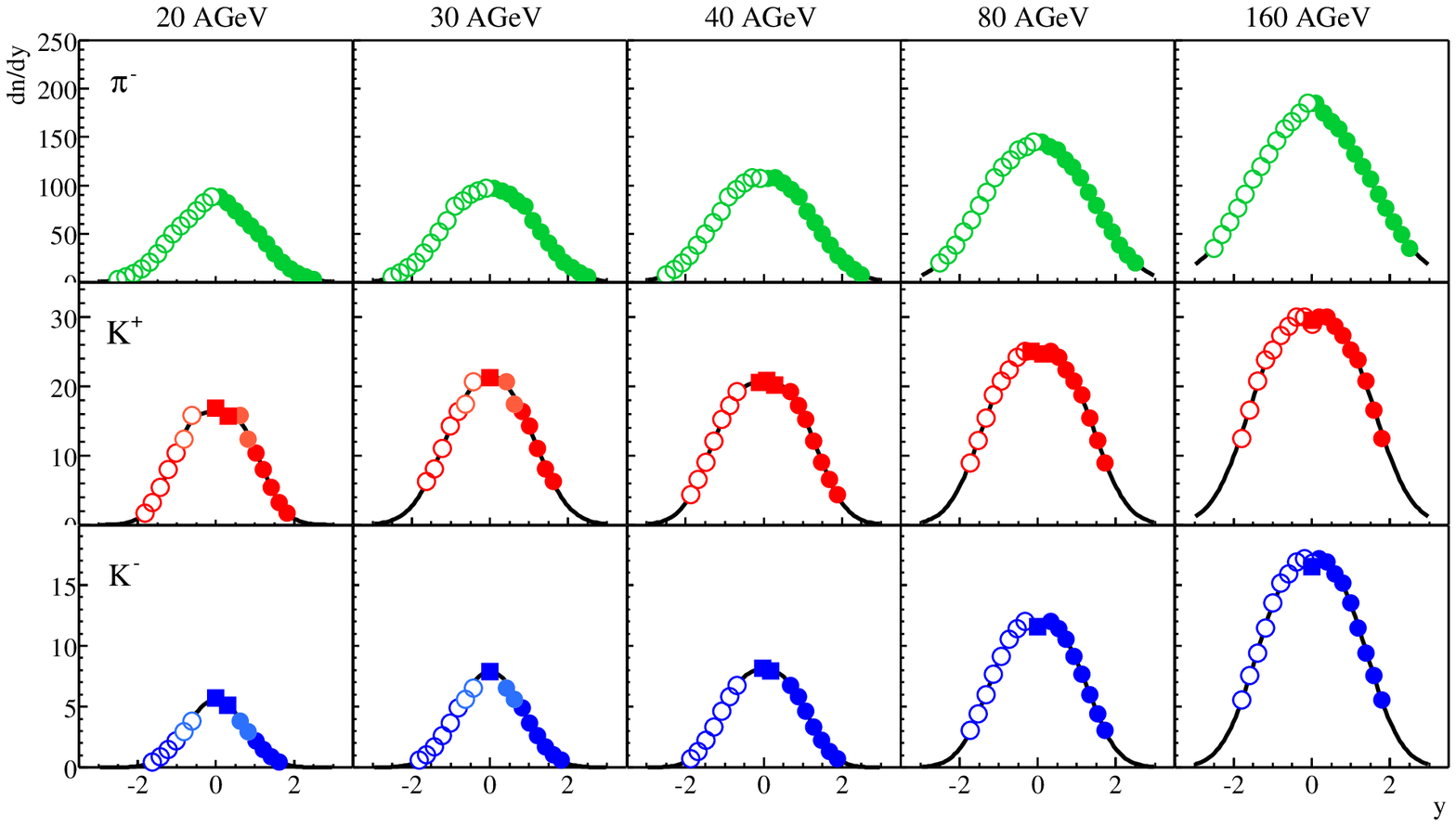,width=0.625\linewidth}
       \epsfig{figure=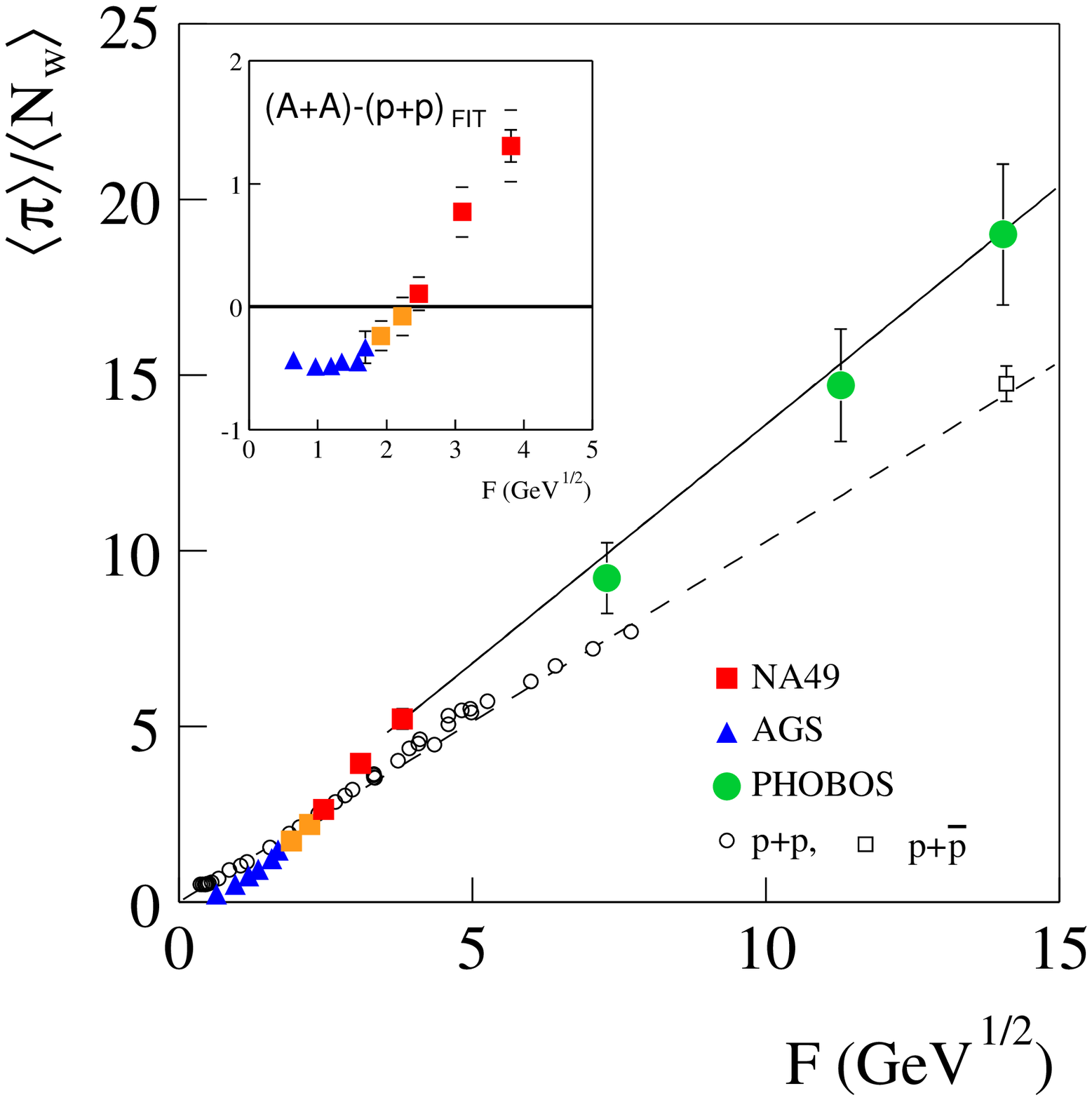,width=0.375\linewidth} }
\end{center}
\caption{Left: Rapidity distributions of $\pi^-$, \kplus\ and \kmin\
       from central Pb-Pb collisions at SPS energies.  The open
       symbols show the data reflected around mid-rapidity. The curves
       are fits to the data with a double Gaussian. Right: Total pion
       multiplicity per wounded nucleon versus the Fermi energy
       measure $F \approx s_{\rm NN}^{1/4}$ for central A-A collisions
       (full symbols) and for p-p interactions (open circles). The
       full line is a linear fit to the A-A data at and above 158
       \agev. The difference between the A-A and p-p data (dashed
       line) is shown in the inset.
\label{fig:rapidity}}
\end{figure}
%------------------------
%
Because the pion momenta are too low for \dedx\ identification their
yields were determined from the negative hadron spectra corrected for
contributions from kaon and weak decays. To determine the total yields
the spectra were extrapolated using the fitted double Gaussians shown
by the curves in the figure.

The energy dependence of the total pion yield (taken to be 1.5 times
the measured charged pion yield) per wounded nucleon (\nw) is shown in
the right-hand plot of \Fi{fig:rapidity} for A-A and p-p(\pbar)
collisions. It is seen that in both colliding systems $\pi/\nw$
increases with energy but that for A-A collisions the rate of increase
becomes larger at the SPS. In the SMES this steepening is explained by
an increase of the effective number of internal degree's of freedom
due to the onset of de-confinement (assumed, in the model, to occur at
about 30~\agev).

The ratio's \kpimin, \kpiplus\ and \lpiavg\ (here $\pi$ denotes the
average of the $\pi^+$ and $\pi^-$ yields) are shown in
\Fi{fig:kpiratio} as a function of collision energy.  The \kpimin\
ratio increases monotonically with energy while the \kpiplus\ ratio
shows a pronounced peak at about 30~\agev; this characteristic energy
dependence seems to be a unique feature of A-A collisions because it
is absent in elementary p-p collisions (open circles in
\Fi{fig:kpiratio}). There appears to be little energy dependence of
the \kpiplus\ ratio between SPS and RHIC. The peak in the energy
dependence of \lpiavg\ is qualitatively understood as a threshold
effect at AGS energies followed by a suppression at larger energies
because the baryo-chemical potential decreases with energy.  The
behavior of \lpiavg\ is reflected in \kpiplus\ by associated
production.

%
%------------------------
\begin{figure}[tbh]
\begin{center}
\mbox{\epsfig{figure=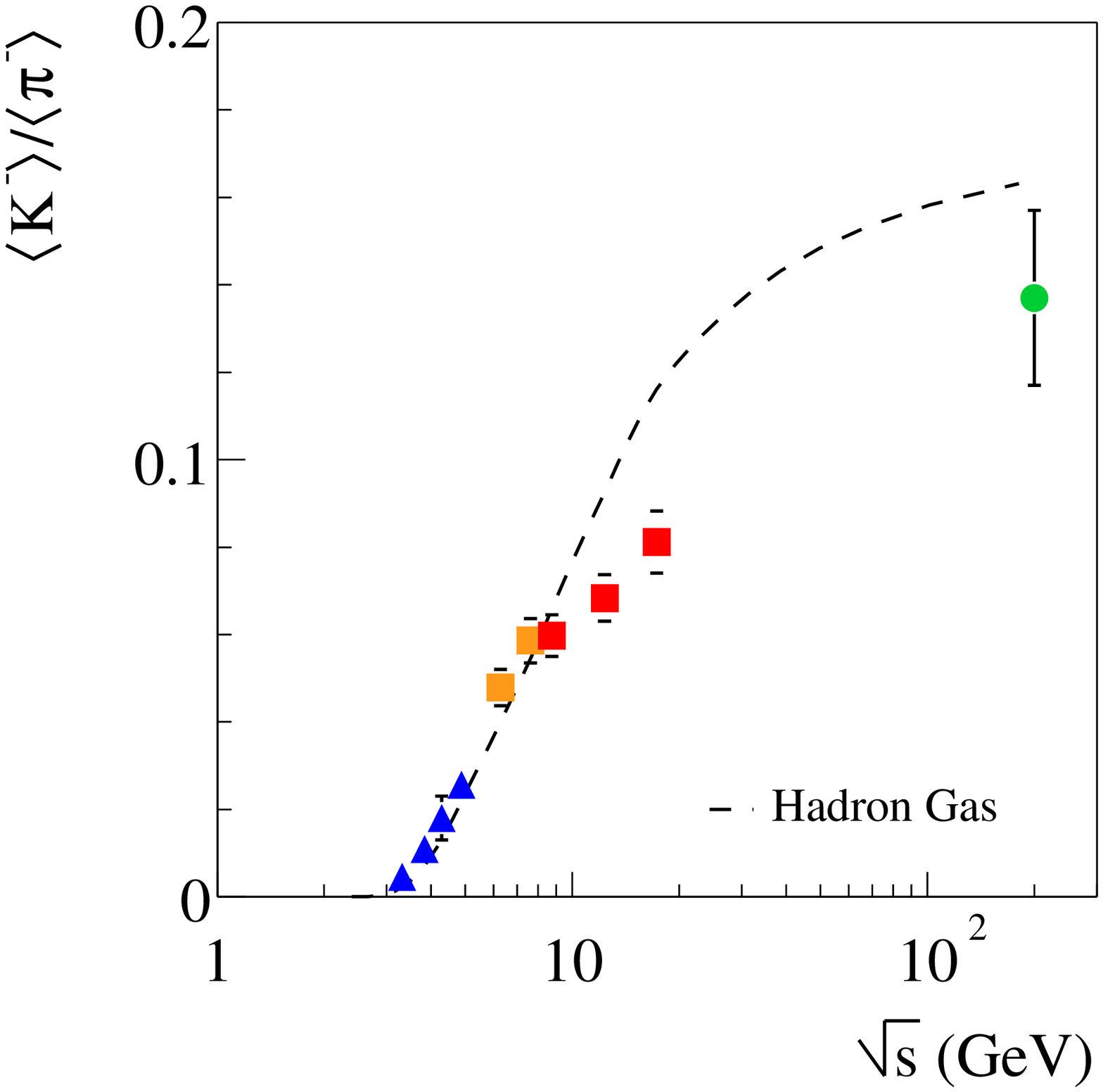,width=0.30\linewidth}
      \epsfig{figure=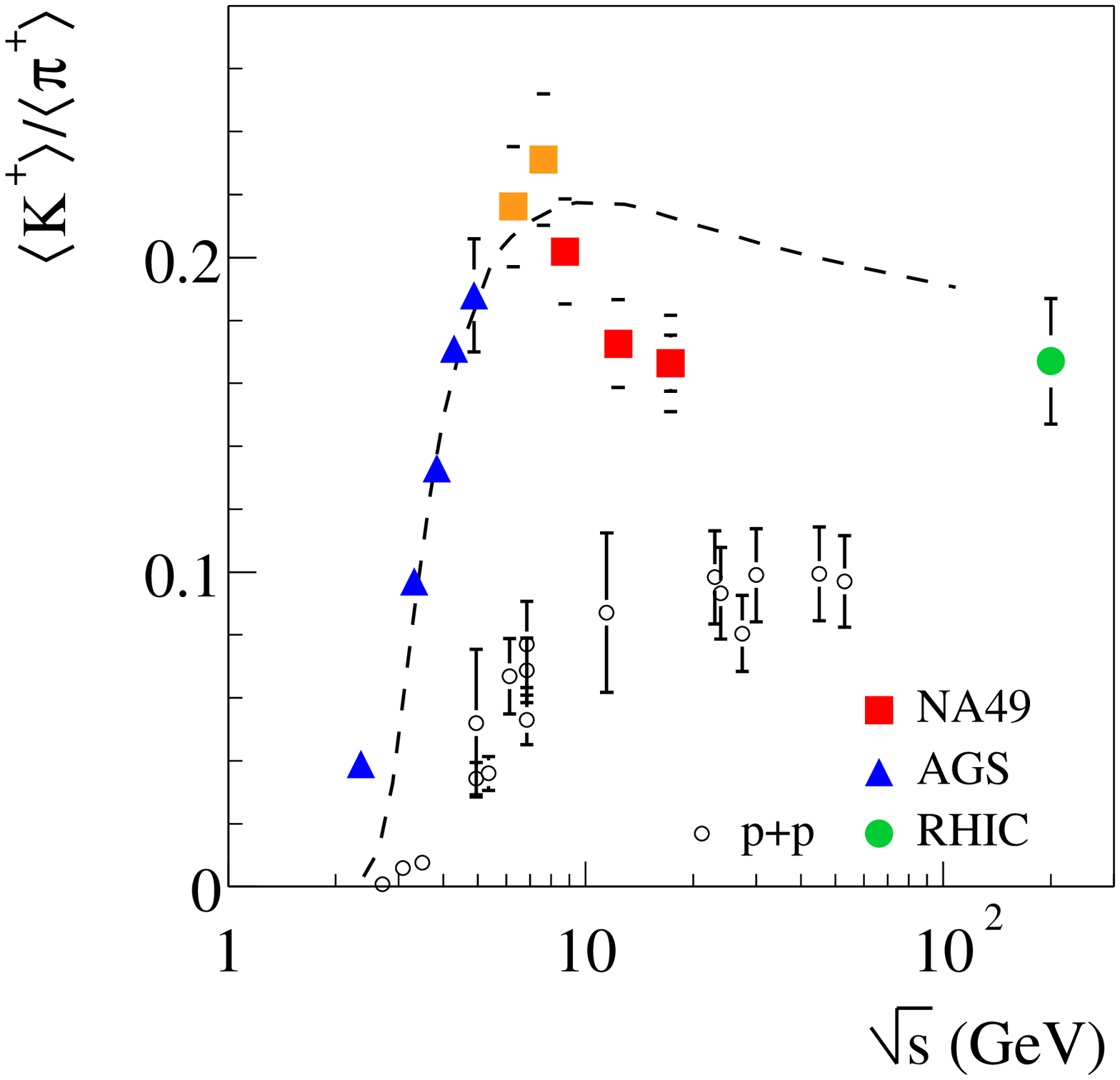,width=0.30\linewidth}
      \epsfig{figure=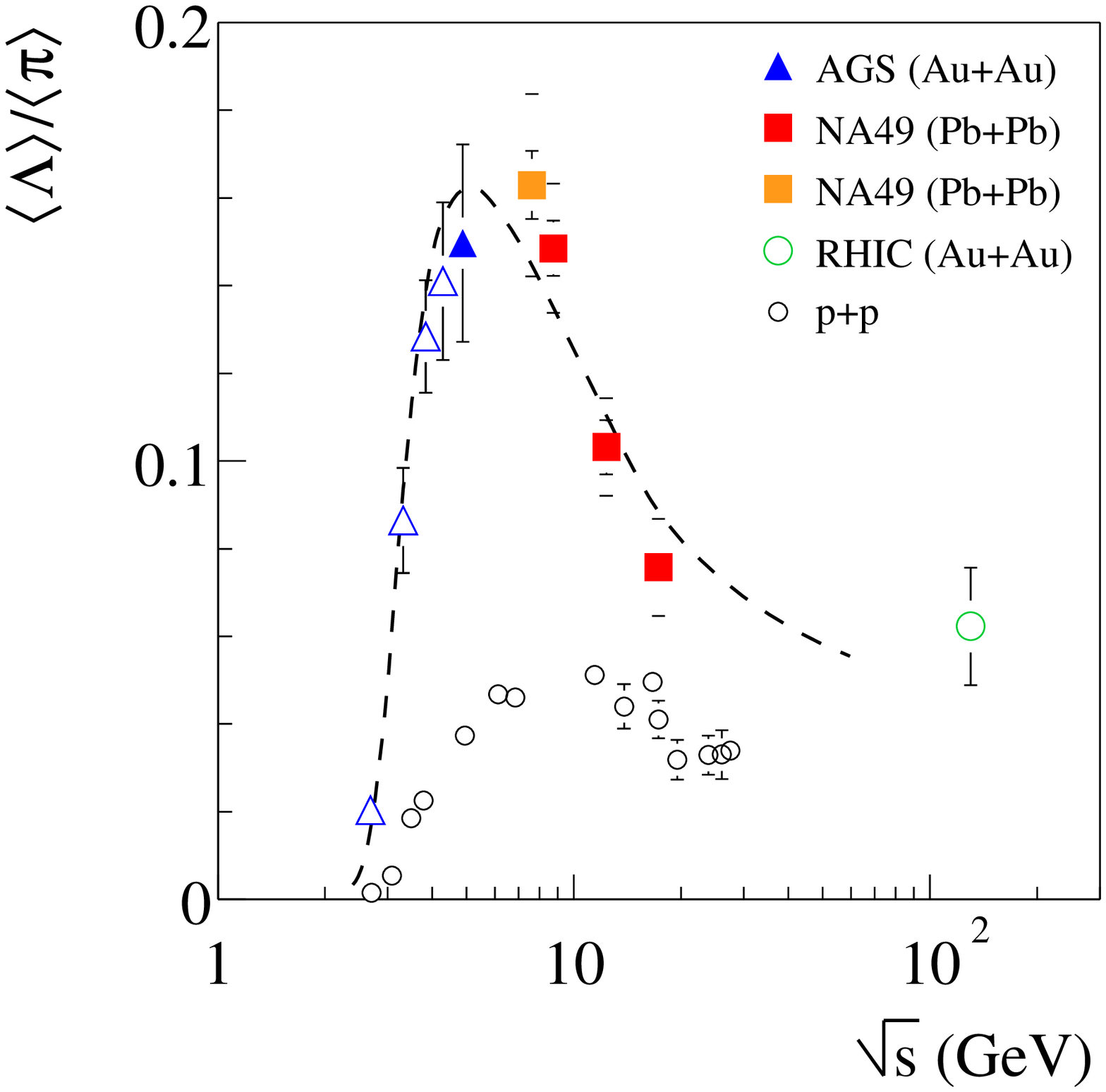,width=0.30\linewidth} }
\end{center}
\caption{Energy dependence of the ratio \kpimin\ (left),
      \kpiplus\ (middle) and \lpiavg\ (right) in central A-A
      collisions and p-p interactions (open circles). The curves are
      predictions from a hadron-gas model.~\protect\cite{ref:krhg}
\label{fig:kpiratio}}
\end{figure}
%------------------------
%

The curves in \Fi{fig:kpiratio} show the prediction from a statistical
hadron gas model.~\cite{ref:krhg} In such models it is assumed that
hadron species are populated according to the available phase
space. Particle ratio's are then described by two parameters only: a
chemical freeze-out temperature \tch\ and a baryo-chemical potential
\mub. The model incorporates an energy dependence by smooth
interpolation of \tch\ and \mub\ obtained from fits to data at
different energies. It is seen from \Fi{fig:kpiratio} that the model
gives a fair description of \lpiavg, captures the trend of \kpimin\
but fails to describe the sharp peak in \kpiplus. Microscopic string
models like UrQMD and HSD also fail to give a satisfactory description
of the data (not shown); for detailed comparisons we refer to a recent
study~\cite{ref:urqmd} by the authors of these models.

Hyperon yields ($\Lambda$, $\Xi$ and $\Omega + \bar{\Omega}$),
normalized to $\pi = 1.5 \tm (\pi^+ + \pi^-)$, are shown as a function
of energy in the left-hand plot of \Fi{fig:smes}.
%
%------------------------
\begin{figure}[tbh]
\begin{center}
\mbox{\epsfig{figure=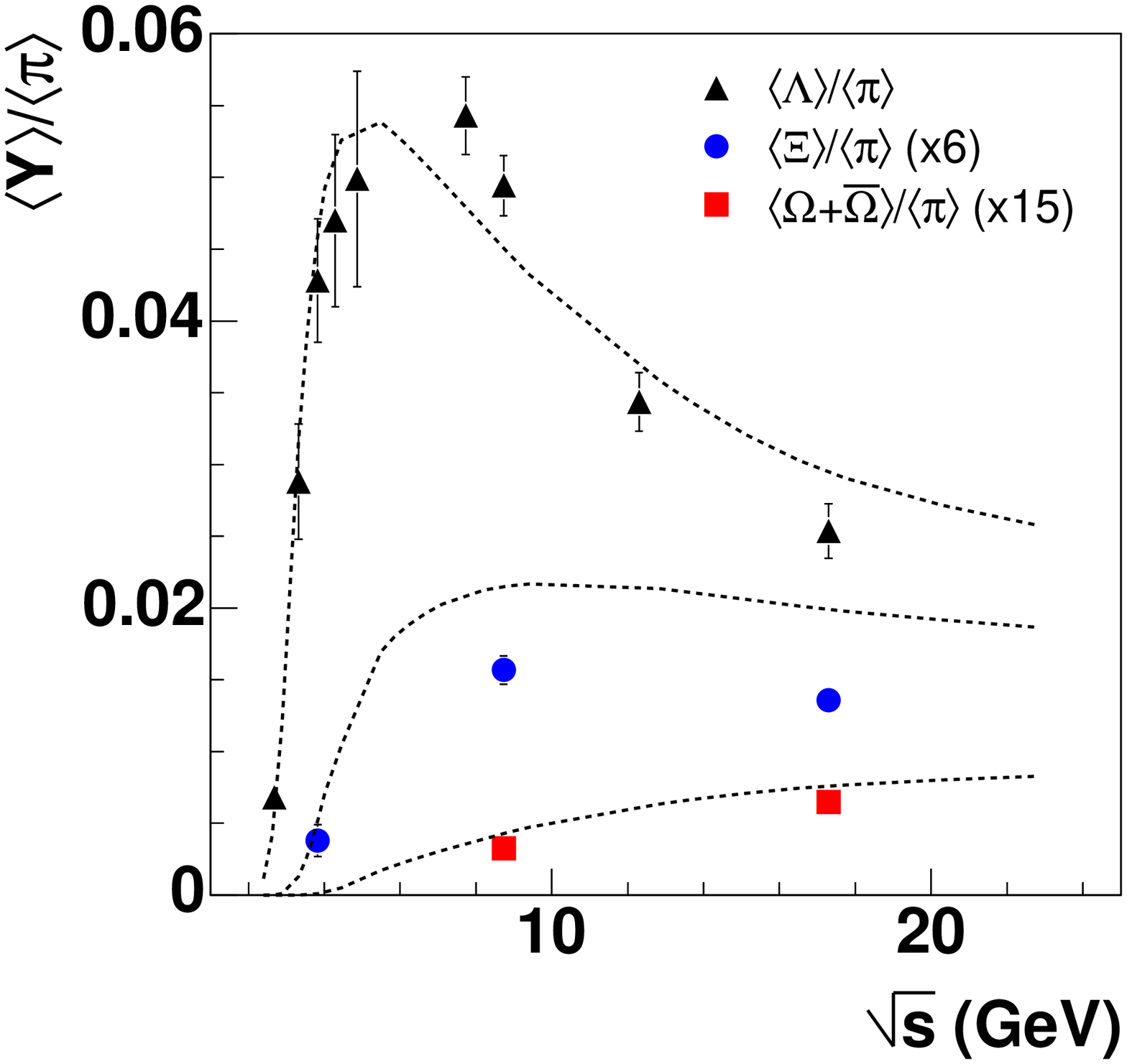,width=0.33\linewidth}
      \hspace{0.1\linewidth}
      \epsfig{figure=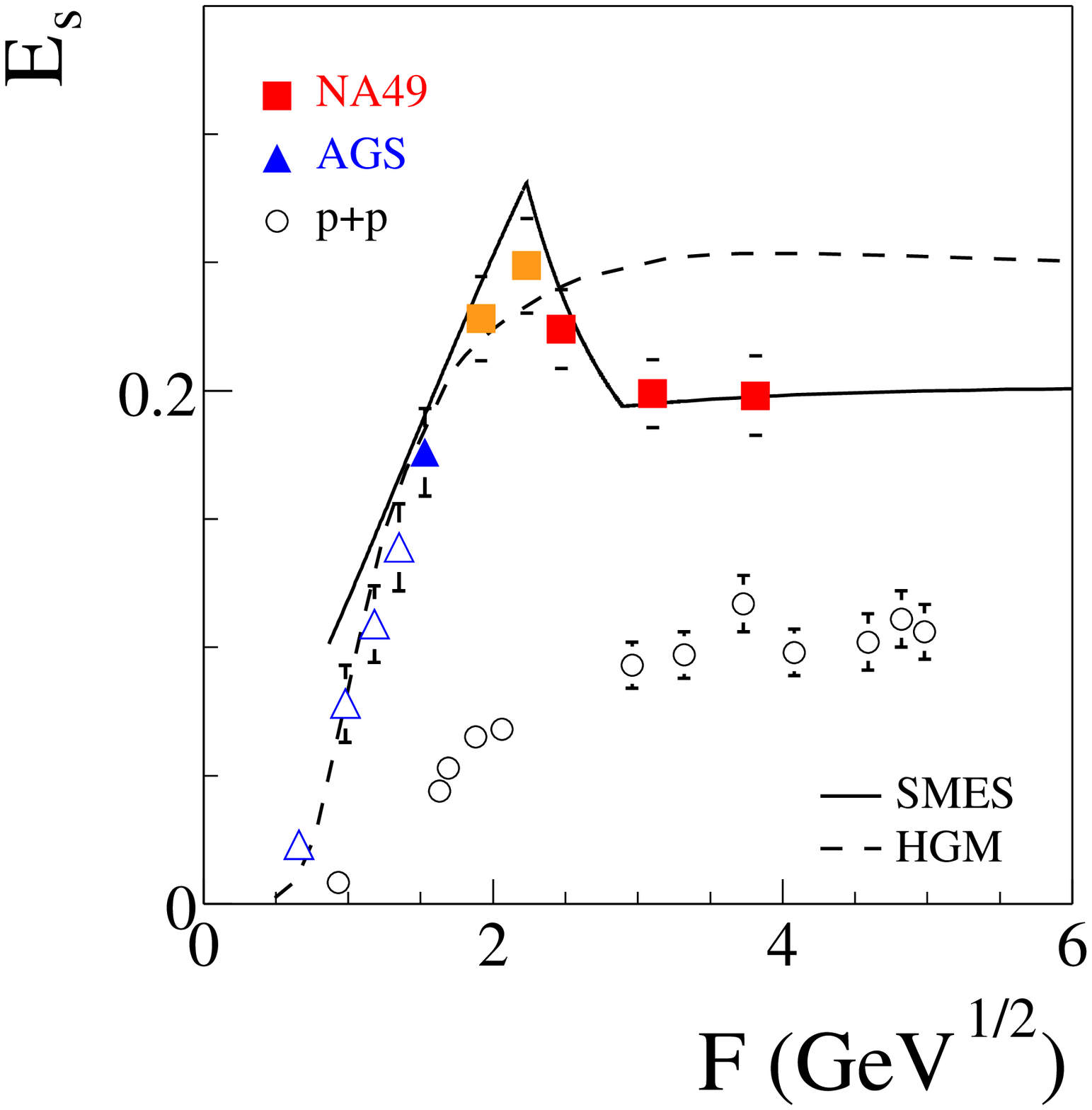,width=0.36\linewidth} }
\end{center}
\caption{Left: Energy dependence of the $\Lambda$, $\Xi$ and
      ($\Omega+ \bar{\Omega}$) to $\pi$ ratio compared to
      predictions from a hadron-gas model.~\protect\cite{ref:pbmhg}
      Right: The strangeness to pion ratio $E_s$ (see text) versus the
      Fermi energy measure $F \approx s_{\rm NN}^{1/4}$ compared to
      predictions from SMES~\protect\cite{ref:smes} (full curve) and a
      hadron gas model~\protect\cite{ref:krhg} (dashed curve, also
      shown in \Fi{fig:kpiratio}).
\label{fig:smes}}
\end{figure}
%------------------------
%
The maximum in $\Lambda / \pi$ (see also \Fi{fig:kpiratio}) is
less pronounced in the ratio $\Xi/\pi$ and seems to be absent
in $\Omega / \pi$. The curves in this plot correspond to the
hadron-gas model of~\cite{ref:pbmhg} which gives a fair description of
the data.

The ratio $E_s = ({\rm K} + \Lambda)/\pi$ versus the Fermi energy
measure $F$ is shown in the right-hand plot of \Fi{fig:smes}. The
full curve shows the prediction from the SMES model which agrees
well with the data. In the SMES $E_s$ is a measure of the strangeness
to entropy ratio and its energy behavior is a direct consequence of the
onset  of de-confinement taking place at about 30~\agev.

%--------------------------------------------------------------------

\section{Summary and outlook}   

The data presented here show that rapid changes of hadron production
properties occur in the SPS energy range. These results can be
understood by assuming that a de-confinement transition takes place at
these energies but it is not clear, at present, if they cannot also be
explained in a purely hadronic scenario. To make progress it is of
interest to measure the energy dependence of particle yields in
collisions of lighter nuclei as well as in elementary p-p and p-A
interactions. Such data would significantly constrain the
models of particle production in heavy-ion collisions.

%--------------------------------------------------------------------

% \section*{Acknowledgments}

% This is where one places acknowledgments for funding bodies etc.
% Note that there are no section numbers for the Acknowledgments, Appendix
% or References.

%--------------------------------------------------------------------

\end{document}